\documentclass[a4paper,11pt]{article}
\usepackage{graphicx}
\usepackage{amsfonts, amsmath, eufrak}

\def\al{\alpha}
\def\be{\beta}
\def\de{\delta}
\def\ga{\gamma}

\def\ep{\epsilon}

\def\te{\theta}
\def\la{\lambda}

\def\si{\sigma}
\def\vp{\varphi}

\def\Ga{\Gamma}

\def\Si{\Sigma}

 \def\calR{{{\cal R}}}

 \def\gotC{{\hbox{$\mathfrak C$}}}
 \def\gotP{{\hbox{$\mathfrak P$}}}

 \def\R{{{\mathbb R}}}

 \def\R{{{\mathbb R}}}

\def\Im{{\hbox{Im}}}

\def\ip{\hbox to4pt{\leaders\hrule height0.3pt\hfill}\vbox to8pt{\leaders\vrule width0.3pt\vfill}\kern 2pt}
\def\del{\partial}
\def\na{\nabla}

\def\arr{\rightarrow}

\def\then{\Rightarrow}

\def\nab#1{{\buildrel #1\over \na}}
\def\ffrac[#1/#2]{\hbox{$\frac{#1}{#2}$}}

\def\({\left(}
\def\){\right)}
\def\[{\left[}
\def\]{\right]}
\def\^#1{{}^{#1}_{\>\cdot}}
\def\_#1{{}_{#1}^{\>\cdot}}
\def\<{\kern -1pt}

\def\[{\begin{equation}}
\def\]{\end{equation}}

\def\ch{\hbox{\rm ch}}
\def\sh{\hbox{\rm sh}}

\begin{document}

\title{Breaking the Conformal Gauge \\ by Fixing Time Protocols}

\author{L. Fatibene$^{1,2}$,  S. Garruto$^{1,2}$, M. Polistina$^3$\\
\\
{\small $^1$Dipartimento di Matematica, Universit\`a di Torino, Italy}\\  
{\small $^2$INFN Sezione Torino- Iniz.~Spec.~QGSKY}\\
{\small $^3$Dipartimento di Fisica, Universit\`a di Torino, Italy}\\
}
 

%\pacs{04.20.Gz, 02.40.Vh, 04.20.q}% PACS, the Physics and Astronomy

\maketitle

\begin{abstract}
We review the definition by Perlick of standard clocks in a Weyl geometry and show how a congruence of clocks can be used to fix the conformal gauge in the EPS framework.
Examples are discussed in details. 
\end{abstract}

\section{Introduction}

In early '70s Ehlers-Pirani-Schild (EPS) introduced an axiomatic approach to geometry of spacetime; see \cite{EPS}. 
Instead of assuming a geometry on spacetime, they assumed as primitive notion the families of the worldlines of particles and of light rays.
They showed that requiring physically reasonable properties for these families one can define  a conformal structure $\gotC$ and a projective structure $\gotP$
on the spacetime $M$.

A {\it conformal structure} is an equivalence class of Lorentzian metrics; $\gotC=[g]$ is made of all the Lorentzian metrics $\tilde g$ conformal to $g$, i.e.~such that there exists
a positive conformal factor $\Phi(x)$ and $\tilde g= \Phi\cdot g$. 
By using a conformal structure $\gotC$ one can define light cones and timelike (spacelike or lightlike) directions. 
On the contrary, distances cannot be defined being rather associated to a specific representative of $g\in \gotC$.

A {\it projective structure} is a class of (torsionless) connections which share the same autoparallel trajectories; $\gotP=[\Ga]$ is made of all connections in the form
$\tilde \Ga^\al_{\be\mu}= \Ga^\al_{\be\mu} - \de^\al_{(\be} V_{\mu)}$ for some covector $V_\mu$.
A projective structure is associated to a free fall; see \cite{geod}.

A projective structure $\gotP$ and a conformal structure $\gotC$ are {\it EPS-compatible} if lightlike geodesics of $\gotC$ are a subset of autoparallel trajectories of the projective structure $\gotP$. In this case one can always choose of representative $\tilde \Ga$ for $\gotP$ in the form
\[
\tilde\Ga^{\al}_{\be\mu}= \{g\}^\al_{\be\mu} -\(g^{\al\ep}g_{\be\mu} - 2\de^\al_{(\be}\de ^\ep_{\mu)}\)A_\ep
\label{EPSConn}\]
for some covector $A$, a representative $g$ of the conformal structure and where $\{g\}$ denotes the Levi-Civita connection of the metric $g$.
Then one also has
\[
\nab{\tilde\Ga}_\la g_{\mu\nu}= -2A_\la g_{\mu\nu}
\]

The triple $(M, \gotC, \tilde \Ga)$ is called a {\it Weyl geometry} where $M$ is a manifold of dimension $m$ (hereafter $m=4$), $\gotC$ is a conformal structure on $M$ and $\tilde\Ga$ is a connection given by  (\ref{EPSConn}) which represents a EPS-compatible projective structure and free fall.
As a matter of fact, the EPS analysis shows that a Weyl geometry is a more natural description of spacetime geometry which contain as a particular case the standard Lorentzian geometry (for $A=0$) though it is more general.

Recently Perlick defined {\it standard clocks} in Weyl geometries; see \cite{Perlick}.
He proved that on any trajectory one con always choose a parametrization (modulo affine transformations) which makes the clock standard.
In section 2 we shall briefly review Perlick's proposal for standard clocks and review the main properties for them.

Then, in section 3, we shall show that any clock is standard for a suitable representative of the conformal structure.
Finally, in section 4, we shall explore some consequences of different conformal fixings induced by different clocks.

\section{Standard Clocks}

In this section we will briefly review the definition of standard clock given by Perlick (see \cite{Perlick}) which is suitable for a Weyl geometry.
A new definition is needed because in a Weyl geometry all definitions are supposed to be local, hence the one usually given (see \cite{WheelerClock}) is not suitable in this context.

\medskip
\noindent \bfseries Definition 1: \mdseries 
A {\it clock} is a timelike curve in $M$, i.e. a $C^{\infty}$ map $\ga : I \rightarrow M: s\mapsto \ga^\mu(s)$, for a  real interval $I\subset \R$, such that its tangent vector $\dot{\ga}^{\mu}(s)$ is timelike for all  $s\in I$. 
\medskip

Two clocks $\ga^{\mu}$ and $\tilde{\ga}^{\mu}$ may share the same worldline ($\Im(\ga) \equiv \Im(\tilde \ga)\subset M$) and differ by the {\it parametrization}: 
this happens if there exists a (orientation preserving) diffeomorphism $\phi:\R\arr \R$ that map one into the other, i.e. if $\ga^{\mu}=\tilde{\ga}^{\mu}\circ\phi$ . Two such clocks are said to be {\it equivalent}. We will also eventually refer to the whole class of equivalent curves $[\ga^{\mu}]$ as a {\it particle} and to the submanifold assicated in $M$ by $\ga^{\mu}$ as {\it worldline} of that particle.\\
When a clock $\ga^{\mu}$ is given is natural to define its 4-velocity and its 4-acceleration by:
\[
v^\mu(s) = \dot{\ga}^{\mu}(s) 
\qquad
a^{\mu}_{(\Ga)}(s)  = \ddot{\ga}^{\mu}(s) + \Gamma^{\mu}_{\alpha \beta}(\ga(s)) \dot{\ga}^{\alpha}(s) \dot{\ga}^{\beta}(s) 
\]
Let us stress that while the 4-velocity depends on the curve only, the 4-acceleration depends on a connection as well.
Hereafter, we shall often use the 4-acceleration associated to the free fall connection $\tilde \Ga$, as well as the one associated to a representative $g\in\gotC$, namely
$a^\mu_{(g)}(s)$.

\medskip
\bfseries Definition 2 \mdseries
A clock $\ga^{\mu}:I \rightarrow M$ is said to be a {\it standard clock} (with respect to a Connection $\Ga$ and a Lorentzian metric $g$) if its 4-acceleration $a^{\mu}_{(\Ga)}(s)$ is $g$-orthogonal  to the 4-velocity $v^{\mu}(s)$, so that:
\[
 g(a_{(\Ga)}, v)=0
\]
The parametrization of a standard clock is usually called a {\it proper time}.

\medskip
Perlick proved that given a Weyl geometry $(\gotC, \tilde \Ga)$ and a conformal representative $g\in \gotC$, then one can always find a parametrization for any particle which corresponds to proper time. 
Consider in fact two equivalent clocks 
\[
\ga^{\mu}(s)=\tilde{\ga}^{\mu}\circ\phi(s)
\qquad\quad
\(s'=\phi(s)\)
\]
Their velocities and accelerations (with respect to any connection $\Ga$) are related by
\[
v^{\lambda}(s)= \dot \phi(s) \>  {\tilde v}^{\lambda}(\phi(s))
\qquad
%\ddot{\ga}^{\lambda}(s)= \dot \phi^2(s) \> \ddot{\tilde{\ga}}^{\lambda}(\phi(s))+\ddot \phi(s)\> \dot{\tilde{\ga}}^{\lambda}(\phi(s)) 
a_{(\Ga)}^{\lambda}(s)=\dot\phi^{2}(s) {\tilde a}_{(\Ga)}^{\lambda}(\phi(s))+ \ddot\phi(s){\tilde{v}}^{\lambda}(\phi(s))
\]
where the dots denote derivatives with respect to $s$ (or $s'$).
From here on we shall understand evaluation at $s$ or $s'$ when it is clear from context.

Then $\tilde{\ga}^{\mu}$ is a standard clock iff we have:
\[
0=g(\tilde{a}_{(\Ga)},\tilde{v})
\quad\then
\ddot\phi= \frac{g( a_{(\Ga)}, v)}{ g(v, v)}\> \dot\phi
\]
which has to be regarded as a differential equation for the unknown function $\phi(s)$.
The solution is unique up to an affine transformation $\phi\mapsto a\phi+b$ with constants $a$, $b\in\R$.

One can also show that this definition of standard clock is consistent with standard General Relativity (GR).
Let us consider a free falling clock in a general Weyl geometry  $(M, [g], \tilde \Ga)$; one has
\[
\begin{aligned}
\frac{d}{ds}&\( g(v,v) \)=
%\frac{d}{ds}(g_{\mu\nu})\dot{x}^{\mu}\dot{x}^{\nu}+2\ddot{x}^{\mu}\dot{x}^{\nu}g_{\mu\nu}=
(\partial_{\lambda}g_{\mu\nu})\dot{\ga}^{\mu}\dot{\ga}^{\nu}
\dot{\ga}^{\lambda}+2\ddot{\ga}^{\mu}\dot{\ga}^{\nu}g_{\mu\nu}=\\
=&
(\nab{\tilde \Ga}_{\lambda}g_{\mu\nu}+2g_{\mu\rho}
\tilde\Gamma^{\rho}_{\nu\lambda})\dot{\ga}^{\mu}\dot{\ga}^{\nu}
\dot{\ga}^{\lambda}+2\ddot{\ga}^{\mu}\dot{\ga}^{\nu} g_{\mu\nu}=\\
=&
2\( g( {a}_{(\tilde\Ga)},v)- A(v) g(v, v) \)
\end{aligned}
\]
with $A(v) = A_\mu v^\mu$.

In the Riemannian case, i.e.~when $\tilde\Ga=\{g\}$, one has:
\[
\frac{d}{ds}\( g(v,v) \)=2 g(a_{(g)}, v)
\label{MetricStandardClock}
\]
which vanishes for standard clocks.
Hence the length of the 4-velocity of a standard clock is constant, which is the definition of proper time in standard GR. 
Then  one can use the affine freedom to fix $|v|^2=-1$.

The notion of standard clock though applies also to non-free fallling clocks. For example, the  clock $\ga:s\mapsto (s, r_0, \te_0, \vp_0)$
is standard in a Schwarzschild metric though it is not free falling. Also in this case of course the tangent vector $\dot \ga$ has constant length
and the clock is parametrized by its proper time.

We already showed in \cite{Fluids} that is we fix a representative $g$ of a conformal structure $\gotC$, then any congruence of timelike curves induces an EPS-compatible 
connection $\tilde\Ga$ for which the timelike curves are autoparallel.

\section{Breaking the Conformal Invariance} 

Hereafter we shall  show that given a congruence of clocks, there is a representative of the conformal structure for which the clocks are standard.
In other words, given a particle there are three relevant objects: a parametrization to make the particle a clock, a representative $g$ of the conformal structure, and a Weyl connection describing free fall. Given two of this objects the third can determined so that the clock is standard with respect to $g$ and $\tilde \Ga$. 

Let us consider first a clock $\ga(s)$ and a representative $\hat g\in\gotC$. In view of equation (\ref{MetricStandardClock}), the clock $\ga$ is standard with respect to the
metric $\hat g$ iff and only if $\hat  g(v,v)$ is constant along the curve $\ga$, which in general is not the case.
However, one can always choose a conformal factor
\[
\Phi(\ga(s))= -\frac{\al^2}{\hat g(v, v)}
\quad\then
g= \Phi\> \hat g
\]
so that $ g(v,v)=-\al^2$ along the curve $\ga$. Then for such a representative $g\in \gotC$ the clock $\ga$ is standard with respect to $g$.
Let us stress that this result is achieved without changing  the clock parametrization.
 
Of course, this procedure does not {\it fix} the representative $g$ of the conformal factor since we have no information about the value of the conformal factor outside the worldline of $\ga$. However, we we have a congruence of clocks which fills (an open set of) the spacetime $M$, then the conformal factor and the metric $g$ will be (locally) fixed uniquely (up to affine transformations) by the requirement of the clocks to be standard.

Once again Weyl geometries show to be less strict than Lorentzian geometries as shown in \cite{Fluids}. We have just shown that considered a representative of the conformal structure $g$, a connection for free fall $\tilde\Ga$ and a clock $\ga$ one can keep fixed two of them and modify the third to obtain a standard clock.

The rest of this section is devoted to show that also a {\it single clock} in fact does the trick of fixing the conformal factor, just because a single clock does in fact allow to define a congruence of clocks. 
This is already contained in EPS and Perlick framework; see \cite{EPS}, \cite{Perlick}. 

One of the EPS axioms prescribes that if one fixes a clock $\ga$, then  for any event $x\in \Im(\ga)$ there exists two neibourhoods $U_x\subset V_x\subset M$
such that for any point $e\in U_x$ there exists two light rays $r_\pm$ through $e$ which intersect the worldline of the clock $\ga$ at two points $p_\pm (e)\in V_x$.
Since $p_\pm\in \Im(\ga)$ they correspond to two values of the parameter along the clock, say $s_\pm (e)$.

Following \cite{Perlick}, one can define two (local) functions $\tau=\frac{1}{2}(s_+ + s_-)$ and $\rho= \frac{1}{2}(s_+ - s_-)$ (which are smooth outside $\Im(\ga)$)
which provide an operational definition of {\it time} and {\it distance} (which  of course depend on the clock).
In particular one can define the hypersurfaces $\Si_t=\{e\in U_x: \tau(e)=t\}$ which are isochronous hypersurfaces. 
If one has enough clocks  $\ga_i$ (one clock is enough when $\dim(M)=2$) the functions $\rho_i$ singles out a point in each $\Si_t$, defining a worldline $P$
which is timelike. Therefore a point $p\in P$ belongs to some $\Si_t$ and one can use $t$ as a parameter along $P$ to define a clock.
In this way one has a congruence of clocks conventionally defined out of the original clock.
Then by requiring that all these clocks are standard one can single out a conventional breaking of the conformal gauge defining a representative $g\in \gotC$
which in turn defines distances and time lapses. 

This construction defines a local chart on spacetime and it is then associated to a special class of observers.
In the next section we shall discuss how one can compute (for the sake of simplicity in 2d special relativity) transition functions between coordinates associated to this kind of observers.

\section{Changing Observers} 

For simplicity let us hereafter restrict to 2-dimensional spacetimes.
Let us  consider $M=\R^2$; let $(x, t)$ be coordinates on $M$ and let particles and light rays be straight lines $\ga_u:\R\arr M: s\mapsto ( us+ x_0, s+ t_0)$, 
with $-1<u<1$ for particles and with $u=\pm1$ for light rays. The coordinates $(x, t)$ chosen $M$ have no special meaning, they are just a mean to write parametrizations of curves.

\subsection{Clocks at rest}

Let us first fix a clock $\ga_{P_0}: s\mapsto (x_0,  t_0+ s)$ through the point $P_0=(x_0, t_0)$  which corresponds to the worldline $x=x_0$. 
For any event $P_1=(x_1, t_1)$ there are always exactly two light rays through $P_1$ which intersect the worldline of the clock at
\[ 
s_\pm(P_1)= {-\( t_0-t_1\)\pm |x_0-x_1|}
\qquad\then
\begin{cases}
\tau= {t_1 -t_0}  \\
\rho=  {  |x_1-x_0|}
\end{cases}
\] 

Accordingly, if the clocks at rest are set to zero on an isochronous line $\tau=\hbox{\it const}$, then they are synchronized.
The two functions  $(\rho, \tau)$ define a coordinate grid and thus defined Cartesian coordinates $(x, t)$ as $|x|=\rho$ and $t=\tau$.
These coordinates, which incidentally coincide with the original coordinates we chose on $M$, are centered at $(x_0, t_0)$, and their origin can be moved anywhere by a translation.

Following EPS prescription an  observer comoving with the clocks defines a function $G_{P_0}$ on $M$ as
\[
G_{P_0}(P_1)=-s_+s_-=-\(t_1-t_0\)^2 + \(x_1-x_0\)^2
\] 
and defines the metric at $P_0$ as the coefficient of the second order expansion of $G_{P_0}$, namely $G_{P_0}(P_1)\simeq g_{\mu\nu}(P_0) (x_1^\mu-x^\mu_0)(x_1^\nu-x^\nu_0)$ which corresponds to the Minkowski metric $g=\eta_{\mu\nu} dx^\mu\otimes dx^\nu=-dt^2+dx^2$ at the point $P_0$.
 
Accordingly,  choosing the congruence of clocks $\ga_{P_0}$ as above corresponds to an inertial observer which sees the clocks at rest, which establishes standard coordinates $(x, t)$ and defines the metric as Minkowski anywhere.
All clocks $\ga_{P_0}$ are standard with respect to $\eta$. For $v=\dot \ga_{P_0}$ one has $g(v,v)=-1$.

\subsection{Clocks at constant speed}

Let us now consider another congruence of clocks $\ga'_{P_0}: s\mapsto (x_0+\be \al s,  t_0+\al s)$  for some velocity $-1<\be<1$ and $\al\in\R^+$.
For any other event $P_1$ one has two light rays through $P_1$ hitting the clock at the parameter
\[
s_\pm = \frac{-(t_0-t_1) + \be(x_0-x_1) \pm | (x_0-x_1) -\be(t_0-t_1) |}{\al(1-\be^2)}
\]
so that it defines the functions
\[
\tau'=  \frac{(t_1-t_0) - \be(x_1-x_0) }{\al(1-\be^2)}  
\qquad\qquad
\rho'=   \frac{| (x_1-x_0) -\be(t_1-t_0) |}{\al(1-\be^2)}
\]

Accordingly, if the clocks are set to zero on an isochronous line $\tau'=\hbox {\it const}$, then they are synchronized.
The functions $(\rho', \tau')$ define a coordinate grid which is associated to Cartesian coordinates $(x', t')$  and they are centered on the clock starting event $(x_0, t_0)$.

Following EPS prescription the  observer comoving with the clocks defines a function $G'_{P_0}$ on $M$ as 
\[
G'_{P_0}(P_1)=-s_+s_-=\frac{-\((t_1-t_0) -\be(x_1-x_0)\)^2 + \((x_1-x_0) -\be(t_1-t_0)\)^2}{\al^2(1-\be^2)^2}
\] 
and defines the metric at $P_0$ as the coefficient of the second order expansion of $G'_{P_0}$, namely $G'_{P_0}(P_1)\simeq g'_{\mu\nu}(P_0) (x_1^\mu-x^\mu_0)(x_1^\nu-x^\nu_0)$ which corresponds to the metric 
\[
g'=g'_{\mu\nu} dx^\mu\otimes dx^\nu=\frac{-dt^2 + dx^2}{\al^2(1-\be^2)}
\]
 at the point $P_0$.
 The velocities of the new clocks are $v'= \al (\del_t +\be\del_x)$ and one has $g'(v', v')= -1$, as for the rest clocks.
 
 %One can exploit the affine invariance of standard parametrizations to set $\al^2(1-\be^2)=1$.
 %This is equivalent to choose {\it identical} clocks in motion and at rest (or fixing the {\it same} units).
 By this choice one has coordinates in the form
 \[
\begin{cases}
t' = \frac{\al }{\al^2(1-\be^2)} \(t - \be x \) \\
x' =  \frac{\al }{\al^2(1-\be^2)} \( -\be t + x \)
\end{cases}
\]

One can easily check that 
\[
g'=\frac{1}{\al^2(1-\be^2)}(-(dt)^2 + (dx)^2)= 
-(dt')^2 + (dx')^2
\]
so that also the new observer defines, in its coordinates $(x',t')$,   the  Minkowski metric.

The phace $\al$ of the clock must be fixed so that $g(\dot \ga'_{P_0}, \dot \ga'_{P_0})=-\al^2(1-\be^2) =-1$, thus $\al=(1-\be^2)^{-1/2}$.
Then the transformation between the two observers is
\[
\begin{cases}
t'=\al(t - \be x) \\
x'=  \al(- \be  t+ x)
\end{cases}
\qquad
\al=\frac{1}{\sqrt{1- \be^2} }
\]
which are known as Lorentz boosts. Accordingly, one can completely recover special relativity from these assumptions.

\subsection{Logarithmic clocks at rest}

Let us now consider an apparently  less trivial example; let us consider a clock at rest which is  parametrized in a non-standard fashion (so that it is not standard for Minkowski metric $g$).
Let us fix a congruence of clocks
\[
\ga''_{Q_0}\colon (-1, +\infty) \to M: \si\mapsto (x_0,  \log(1+\si))
\]
and fix a value $\si_0$ of the parameter so that $1+\si_0=e^{t_0}$.
One can use an affine reparametrization to define a clock starting at $(x_0, t_0)$, namely
\[
\ga''_{P_0}\colon s\mapsto  (x_0,  \log(1+\si_0+s))
\]

Being this a reparametrization of the clock at rest we know in general that the observer associated to this clock will define a metric $g$ 
which is conformal to the metric associated to the clock at rest (i.e.~the Minkowski metric $g= -dt^2 +dx^2$). In other words we expect $g''= \Phi\cdot g$.
By considering light rays through an event $P_1$ one has intersections with the clock at
\[
s_\pm= e^{t_1\pm |x_1-x_0|}- e^{t_0}
\]
and define the two functions
\[
\tau''= e^{t_1} \ch(x_1-x_0)-e^{t_0}
\qquad\qquad
\rho''=e^{t_1} \sh|x_1-x_0|
\]
Notice that this time the coordinate grid is curvilinear (i.e.~the lines of the grid are not geodesics of $g$).

For the metric at $P_0$ we obtain the function 
\[
G''_{P_0}(P_1)= - \(e^{t_1+ |x_1-x_0|}- e^{t_0}\)\(e^{t_1- |x_1-x_0|}- e^{t_0}\)
\]
Expanding it to second order one gets the metric
\[
g''=e^{2t}\(-dt^2+dx^2\)
\label{stclock}
\]
which is as expected conformal to the Minkowski metric $g$. 
The clocks are standard with respect to the metric $g''$.
Thus as a first result one sees explicitly that the choice of the congruence of clocks does in fact fixes the conformal gauge.

The conformal factor is clearly non-constant. However, one can find regions in which the conformal factor is approximately constant.
For example, around an event at $t_0=10$, in the region $t\in [9.98995, 10.00995]$ the conformal factor does change by less than $1\%$.
(We do not have to stress that here scales are completely arbitrary and one can adjust constants in the example so that this interval could span thousands of years.)

%An observer living in that region with instruments with an precision of $1\%$ does not appreciate any curvature of spacetime.
%et us also stress that the absolute value of the conformal factor would be irrelevant. The observer in fact would define its unit so that its metric is exactly Minkowskian at time $t=10$ and sees no deviation within the region where the conformal factor is approximately constant.

%Only the change of the conformal factor is relevant and one does not need to justify why conformal factor today is approximately $1$.
%We defined unit 3 century ago and we are fine until our instruments cannot appreciate a change in the conformal factor at the scales of centuries.

An observer which is comoving with the congruence of clocks does also define new coordinates
\[
\begin{cases}
t'= e^t\ch(x) -1\\
x'= e^t\sh(x)\\
\end{cases}
\label{coord}
\]
In these new coordinates the metric reads as
\[
g''= -dt'^2 + dx'^2
\]
In fact one can easily check that the metric (\ref{stclock}) was itself flat, beside being conformal to (another) flat metric $g$.
These coordinates do not cover the whole Minkowski, due to the singularities of the clocks.
In fact what we have is two different Minkowski spaces and a region in the first Minkowski is conformally mapped into a region of the other.
The map does not preserve the metric, but being the image metric flat it can be written as Minkowski in another coordinate system.

\begin{figure}[h] %  figure placement: here, top, bottom, or page
   \centering
   \includegraphics[width=5cm]{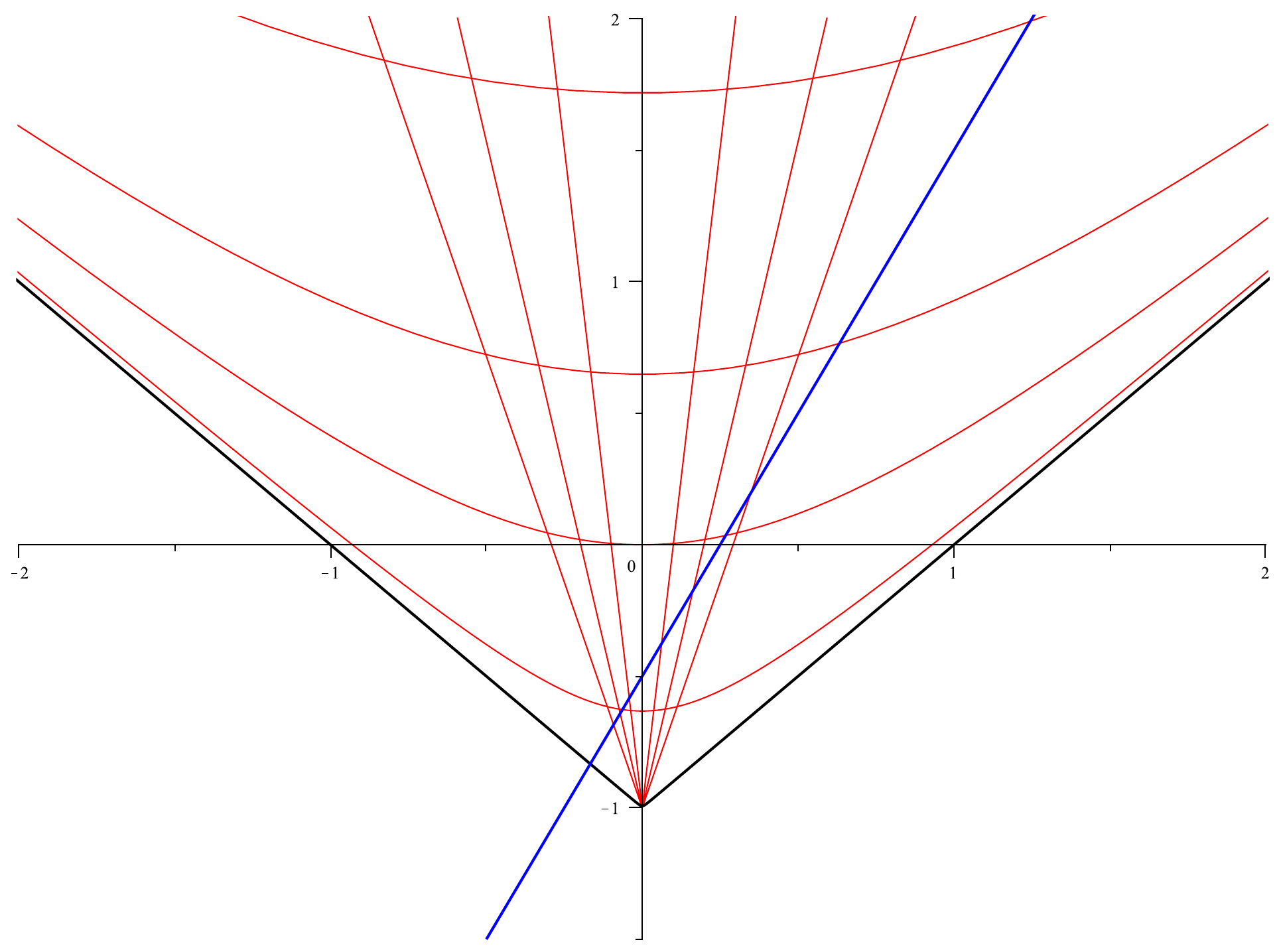} 
   \caption{\small\it Grid for coordinates $(x', t')$ with a free falling particles passing through $P=(x=0.5 , t=0.5)$
   with respect to original coordinates $(x, t)$.}
   \label{fig1}
\end{figure}

An observer living in the new copy of Minkowski can use coordinates  (\ref{coord}) and 
see a spacetime which is exacly Minkowski in a Cartesian coordinate system. Such an observer is entitled as the original one to define special relativity.
However, both observer are looking at the same spacetime and in that spacetime free falling particles either falls along geodesics of $g$ or along geodesics of $g''$. 

If we decide they fall along geodesics of $g$
\[
P\colon s\mapsto (x_0+u s, t_0+s)
\]
for some $-1<u<1$ and in the coordinates $(x, t)$
then such worldlines will be curved (i.e.~non-geodesics) for $g''$
 and  they read in the observer coordinates $(x', t')$ as
\[
P\colon s\mapsto (e^{t_0+s}\sh(x_0+u s), e^{t_0+s}\ch(x_0+u s) -1) 
\]

This of course is not a geodesics of $g''$ and the observer would see it accelerating as soon as it exit the safe region in which the conformal factor is approximately constant.

\begin{figure}[h] %  figure placement: here, top, bottom, or page
   \centering
   \includegraphics[width=5cm]{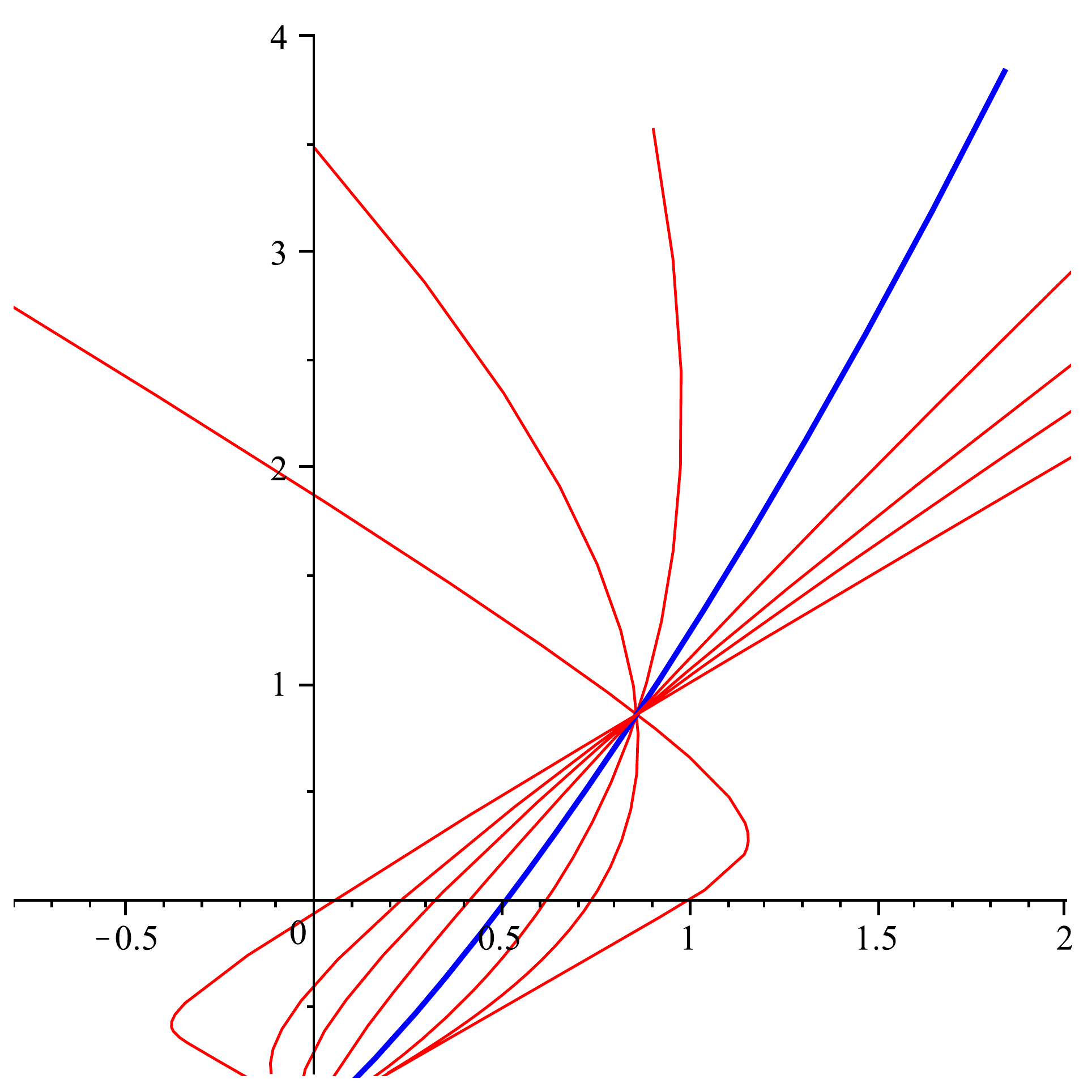} 
   \caption{\small\it  Free falling particles passing through $P=(x=0.5 , t=0.5)\simeq(x'=0.85, t'=0.85)$ with respect to coordinates $(x', t')$}
   \label{fig1}
\end{figure}

The observer would judge that forces are acting on the particle to deflecting it. It would see the force acting universally on any particle in the same way, so it would probably judge the force to be gravitational.
However, in its spacetime there is no gravitational source (since we started off from Minkowski spacetime). 
Accordingly, the observer would judge that its spacetime is filled with some mysterious dark source which affects the motion of objects at non-local scales while it compensates at local scales.

 This is a toy model, and of course we are not claiming that it is a realistic explanation of dark sources.
Still it is based on a somehow simple assumption (atomic clocks are not standard on long times) which we think should not be dismissed on an {\it a priori} basis but should be tested and proven false.

\subsection{Quadratic clocks at rest}

In the previous examples we chose non-standard clocks which eventually turned out to be standard for another Minkowski metric.
However, that was due to the specific form of clocks we considered. Essentially, any other choice gives a new metric which is not flat anymore.

Let us now try and fix a congruence of non-standard clocks:
\[
\ga'''_{Q_0} \colon [0, +\infty)\arr M: \si\mapsto (x_0,  \si^2)
\]
and fix a value $\si_0$ of the parameter so that $\si_0=\sqrt{t_0}$.
One can use an affine reparametrization to define a clock starting at $(x_0, t_0)$, namely
\[
\ga'''_{P_0}\colon s\mapsto  (x_0,  (\si_0+s)^2)
\]

Being this a reparametrization of the clock at rest we know in general that the observer associated to this clock will define a metric $g'''$ 
which is conformal to the metric associated to the clock at rest (i.e.~the Minkowski metric $g= -dt^2 +dx^2$). In other words we expect $g'''= \Phi\cdot g$.
By considering light rays through an event $P_1$ one has intersections with the clock at
\[
s_\pm= \sqrt{t_1\pm |x_1-x_0|}- \sqrt{t_0}
\]
and define the two functions
\[
\begin{aligned}
\tau'''=& \hbox{$\frac{1}{2}$}\(\sqrt{t_1+ |x_1-x_0|}+ \sqrt{t_1- |x_1-x_0|}\) -\sqrt{t_0} \\
\rho'''=& \hbox{$\frac{1}{2}$}\(\sqrt{t_1+ |x_1-x_0|}- \sqrt{t_1- |x_1-x_0|}\) 
\end{aligned}
\]
Notice that again the coordinate grid is curvilinear.

For the metric at $P_0$ we obtain the function 
\[
G'''_{P_0}(P_1)= - \(\sqrt{t_1+ |x_1-x_0|}- \sqrt{t_0}\)\(\sqrt{t_1- |x_1-x_0|}- \sqrt{t_0}\)
\]
Expanding it to second order one gets the metric
\[
g'''=\frac{1}{4t}\(-dt^2+dx^2\)
\label{stclock1}
\]
which is as expected conformal to the Minkowski metric, though this time $g'''$ is not flat. 
The clocks are standard with respect to the metric $g'''$.
Again the choice of the congruence of clocks does in fact fixes the conformal gauge.

The conformal factor is clearly non-constant. However, one can find regions in which the conformal factor is approximately constant.
For example, around an event at $t_0=1$, in the region $t\in [0.99, 10.1]$ the conformal factor does change by less than $1\%$.

An observer living in that region with instruments with an precision of $1\%$ does not appreciate any curvature of spacetime.
Let us also stress that the absolute value of the conformal factor would be irrelevant. The observer in fact would define its unit so that its metric is exactly Minkowskian at time $t=1$ and sees no deviation within the region where the conformal factor is approximately constant.

Only the change of the conformal factor is relevant and one does not need to justify why conformal factor today is approximately $1$.
We defined unit 3 century ago and we are fine until our instruments cannot appreciate a change in the conformal factor at the scales of centuries.

An observer which is comoving with the congruence of clocks does also define new coordinates
\[
\begin{cases}
t'=\frac{1}{2}\(\sqrt{t+x} +\sqrt{t-x} \) \\
x'=\frac{1}{2}\(\sqrt{t+x} -\sqrt{t-x} \) \\
\end{cases}
\label{coord1}
\]
In these new coordinates the metric reads as
\[
g'''= 2\frac{t'^2-x'^2}{t'^2 + x'^2}\( -dt'^2 + dx'^2\)
\]

These coordinates do not cover the whole Minkowski, due to the singularities of the clocks.

\begin{figure}[h] %  figure placement: here, top, bottom, or page
   \centering
   \includegraphics[width=5cm]{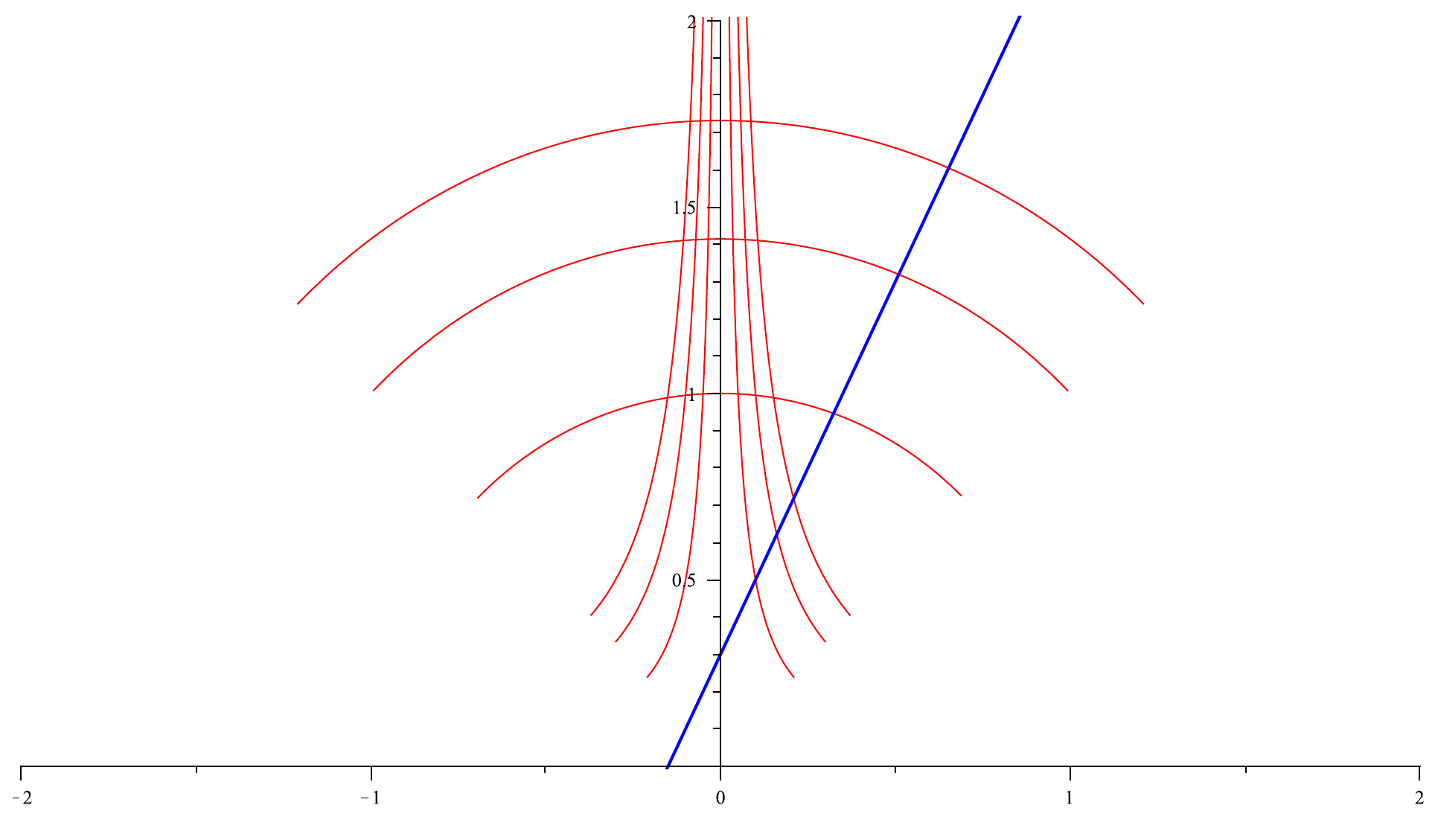} 
   \caption{\small\it Grid for coordinates $(x', t')$ with a free falling particles passing through $P=(x=0.1 , t=0.5)$
   with respect to original coordinates $(x, t)$.}
   \label{fig1}
\end{figure}

However, an observer living in the covered region, quite far away from the boundary,  can use them and 
if it lives in a small region where the conformal factor is approximately constant consider them as approximately Cartesian.

However, if the observer believes that its clock is standard with respect to Minkowski metric it would regard its coordinates as being exactly Cartesian and free falling particles to follow straight lines in spacetime. In fact free falling particle would fall the line 
\[
P \colon s\mapsto (x_0+u s, t_0+s)
\]
for some $-1<u<1$ and
in the coordinates $(x, t)$ which read in the observer coordinates $(x', t')$ as
\[
P \colon s\mapsto \(
\begin{aligned}
\hbox{$\frac{1}{2}$}\(\sqrt{t_0+s+x_0+u s} -\sqrt{t_0+s-x_0-u s} \) \\
\hbox{$\frac{1}{2}$}\(\sqrt{t_0+s+x_0+u s} +\sqrt{t_0+s-x_0-u s} \)
\end{aligned}
\) 
\]
This of course is not a geodesics of $g'''$ and the observer would see it accelerating as soon as it exit the safe region in which the conformal factor is approximately constant.

\begin{figure}[h] %  figure placement: here, top, bottom, or page
   \centering
   \includegraphics[width=5cm]{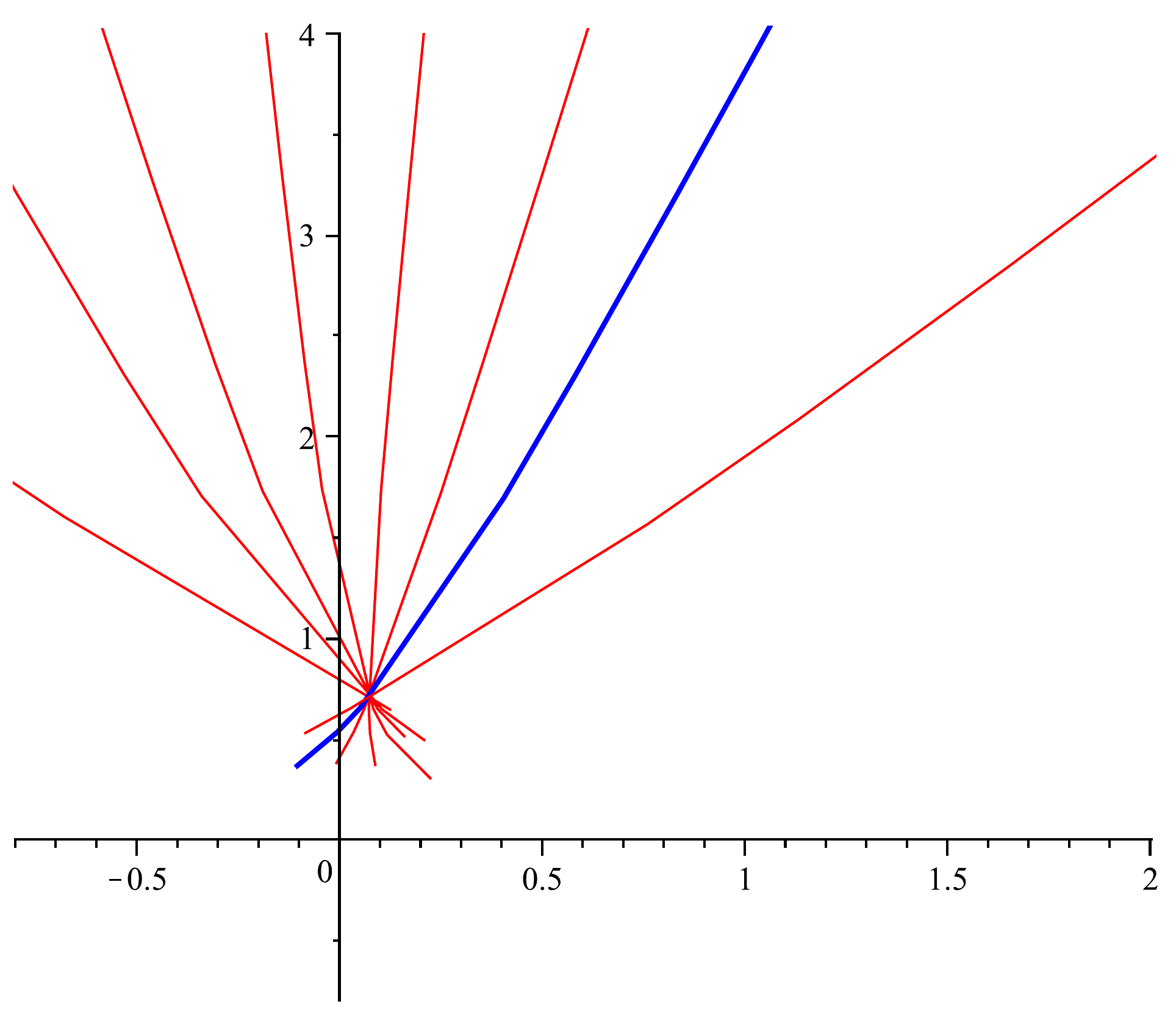} 
   \caption{\small\it  Free falling particles passing through $P=(x=0.1 , t=0.5)\simeq(x'=0.07, t'=0.70)$ with respect to coordinates $(x', t')$}
   \label{fig1}
\end{figure}

\section{Conclusions} 

Here we showed how one can fix the conformal gauge by considering a clock (or a congruence of clocks).
We also investigate the effects of erroneously regarding non-standard clocks for being standard.
Here we consider the effect at a kinematical level as already back to EPS it was noted that (see \cite{EPS}):
{\quote\small\it This last step {\rm [imposing the Riemannian axiom]} seems unavoidable on empirical grounds if equality 
of gravitational time (as given by Weyl arc length and measurable, for example, by the method of Kundt and Hoffmann)
and atomic time is assumed. This is because the latter is transported in an integrable fashion, 
which was pointed out by Einstein in his criticism of Weyl's theory and supported by (among other things)
the consistency of the interpretations of observed red-shifts. (But how compelling is time-equality postulate?)

}

\medskip

We agree that there is no fundamental or real reason to accept {\it a priori} that atomic clocks are standard and that this is something to be experimentally tested.

We also have to notice that there are dynamics in which such effects naturally arise and are dynamically controlled.
Let us consider a Palatini $f(\calR)$-theory of gravitation (see \cite{f(R)}) in which one has two metrics, an original metric $g$ and a conformal metric $\tilde g$
by a conformal factor $f'(\calR)$ which on-shell can be seen to be a function of the trace of matter stress tensor.
For example in cosmology the conformal factor is expected to be a function of cosmological time.

When we provide an operational definition for atomic clocks we of course cannot {\it assume} to be standard with respect to $g$ or $\tilde g$, especially to the precision needed to guarantee it to be standard up to cosmological scales, i.e.~for time scales up to billion years, nor we are able to test it at these scales.
Thus we can speculate about what would happen if the atomic clocks were {\it not} standard with respect to $g$ or $\tilde g$.

If atomic clocks were standard with respect to $\tilde g$ then we could perform a Legendre transformation and obtain standard GR (as already noticed in EPS analysis; see also \cite{EG}).
However, if atomic clocks were {\it not} standard with respect to $\tilde g$ then we showed that it would be standard with respect to a suitable metric $g$ conformal  to $\tilde g$.
This would allow to extend EPS interpretation to the metric sector, assuming that atomic clocks (and then rulers and scales; see \cite{Kepler}) are standard with respect to the original $g$. This will eventually lead (at least) to a wrong estimation of cosmological distances with respect to standard GR cosmology. 
This effect being in addition to the effective sources which are typical of extended theories of gravitation which are used to model dark sources; see \cite{C1}.

\end{document}